\documentstyle[12pt,margins,psfig]{article}

\def\MeV{\,{\rm MeV}}

\def\cmm2{{\,\rm cm^{-2}}}
\def\cm2{{\,{\rm cm}^2}}
\def\cmm3{{\,{\rm cm}^{-3}}}
\def\gcmm3{{\,{\rm g\,cm^{-3}}}}

\def\la{\mathrel{\mathpalette\fun <}}

\def\fun#1#2{\lower3.6pt\vbox{\baselineskip0pt\lineskip.9pt

\ialign{$\mathsurround=0pt#1\hfil##\hfil$\crcr#2\crcr\sim\crcr}}}

\begin{document}
\baselineskip=16pt
\pagestyle{empty}
\begin{center}
\rightline{FERMILAB--PUB-97/035-A}
\rightline{astro-ph/9703065}
\rightline{submitted to {\it Physical Review D}}
\vspace{.2in}

{\bf EFFECT OF FINITE NUCLEON MASS ON PRIMORDIAL NUCLEOSYNTHESIS}\\
\vspace{.2in}

Robert E. Lopez,$^{1}$ Michael S. Turner,$^{1,2,3}$ and Geza Gyuk$^{4}$  \\
\vspace{.1in}

$^1${\it Department of Physics,\\
The University of Chicago, Chicago, IL  60637-1433}\\
$^2${\it NASA/Fermilab Astrophysics Center, \\
Fermi National Accelerator Laboratory, Batavia, IL  60510-0500}\\
$^3${\it  Department of Astronomy \& Astrophysics,\\
Enrico Fermi Institute,
The University of Chicago, Chicago, IL 60637-1433}\\
$^4${\it International School for Advanced Studies (SISSA) \\
Via Beirut 4, 34014 Trieste, Italy}
\end{center}

\vspace{.3in}

\centerline{\bf ABSTRACT}
\bigskip

\noindent 
By numerically evaluating five-dimensional phase-space integrals, we have
calculated the small effect of finite nucleon mass on the weak-interaction
rates that interconvert protons and neutrons in the early Universe. We have
modified the standard code for primordial nucleosynthesis to include these
corrections and find a small, systematic increase in the $^4$He yield, $\Delta
Y/Y \simeq (0.47-0.50)\% $, depending slightly on the value of
the baryon-to-photon ratio $\eta$. The fractional changes in the abundances of
$\rm D$, $^3{\rm He}$, and $^7{\rm Li}$ range from $0.06\%$ to $2.8\%$ for
$10^{-11} \leq \eta \leq 10^{-8}$.

\newpage
\pagestyle{plain}
\setcounter{page}{1}
\section{Introduction}

Primordial nucleosynthesis is one of the cornerstones of the hot big-bang
cosmology.  The consistency between its predictions for the abundances of D,
$^3$He, $^4$He and $^7$Li and their inferred primordial abundances provides
its earliest test. Further, big-bang nucleosynthesis has been used to obtain
the best determination of the baryon density \cite{Copi95,Yang84,Walker91} and
to test particle-physics theories, e.g., the stringent limit to the number of
light neutrino species \cite{Copi97,Schvartsman69,Steigman88}.

Scrutiny of primordial nucleosynthesis, both on the theoretical side and on
the observational side, has been constant: Reaction rates have been updated
and their uncertainties quantified \cite{Krauss90,Smith93,Kernan93};
finite-temperature corrections have been taken into account \cite{Dicus82};
the effect of inhomogeneities in the baryon density explored
\cite{Matthews96}; the slight effect of the heating of neutrinos by $e^\pm$
annihilations has been computed \cite{Dodelson92,Fields93}; the primordial
abundance of $^7$Li has been put on a firm basis and its destruction in stars
has been studied
\cite{Spite82,Spite84,Rebolo88,Hobbs91,Olive92,Thorburn94,Pinson92,Lemoine96};
the production and destruction of $^3$He and the destruction of D have been
studied \cite{Yang84,Dearborn86,Steigman95,Copi95b}; and
astrophysicists now argue about the third significant figure in the primordial
$^4$He abundance \cite{Pagel91,Pagel92,Olive95}.

A measure of the progress in this endeavour is provided by the shrinking of
the ``concordance region'' of parameter space.  The predicted and measured
primordial abundances agree provided: the baryon-to-photon ratio lies in the
narrow interval $2 \times 10^{-10}\le \eta \le 7\times 10^{-10}$ and the
equivalent number of light neutrino species $N_\nu \le 4$
\cite{Copi95,Walker91}. The shrinking of the concordance interval motivates
the study of smaller and smaller effects. In particular, once the primeval
deuterium abundance is accurately determined in high-redshift hydrogen clouds,
the baryon-to-photon ratio will be pegged to an accuracy of order 10\%,
and recent progress suggests that this will happen sooner rather than
later~\cite{Tytler97}. In turn, this will reduce the uncertainty in the
predicted $^4$He abundance due to the baryon density to $\Delta Y \simeq \pm
0.005$.

The weak-interaction rates govern the neutron-to-proton ratio and thereby are
crucial to the outcome of nucleosynthesis; e.g., the mass fraction of $^4$He
produced is roughly twice the neutron fraction at the time nucleosynthesis
commences ($T\sim 0.08\MeV$). In the standard code~\cite{Kawano92,Kawano89}
these rates are computed in the infinite-nucleon-mass limit because
this simplifies
the expressions for the rates to one-dimensional integrals.  The finite-mass
corrections involve terms of order $m_e/m_N$, $T/m_N$, and $Q/m_N$, which are
all of the order of 0.1\%. Here $m_e$ is the electron mass, $m_N$ is the
nucleon mass, $T\sim {\cal O}(\MeV )$ is the temperature during the epoch of
nucleosynthesis, and $Q=m_n-m_p=1.293\MeV$ is the neutron-proton mass
difference. As it turns out, the corrections to the rates are actually of
order a few percent, so that the change in $^4$He abundance is expected to be
a few parts in the third significant figure ($\Delta Y/Y \sim
-0.5 \delta\,$rate/rate).

Because the third significant figure of the primordial $^4$He abundance is now
relevant, we set out to calculate the finite-nucleon-mass corrections to
the weak-interaction rates by numerically integrating the exact expressions
for the rates. This involved accurately ($\la 0.1\,\%$) evaluating five-dimensional
rate integrals.  We incorporated our results into the standard
nucleosynthesis code and found that $\Delta Y/Y \approx (0.47-0.50)\%$,
depending on the value of $\eta$. In the next Section we discuss the weak
rates, and in the following Section finite-mass corrections to the weak
rates. In Sec.~IV we discuss our changes to the standard nucleosynthesis code,
and we finish with a discussion of our results for the changes in the yields
of $^4$He and the other light elements.

\section{Weak-interaction Rates}

\begin{figure}[t]
\centerline{\psfig{figure=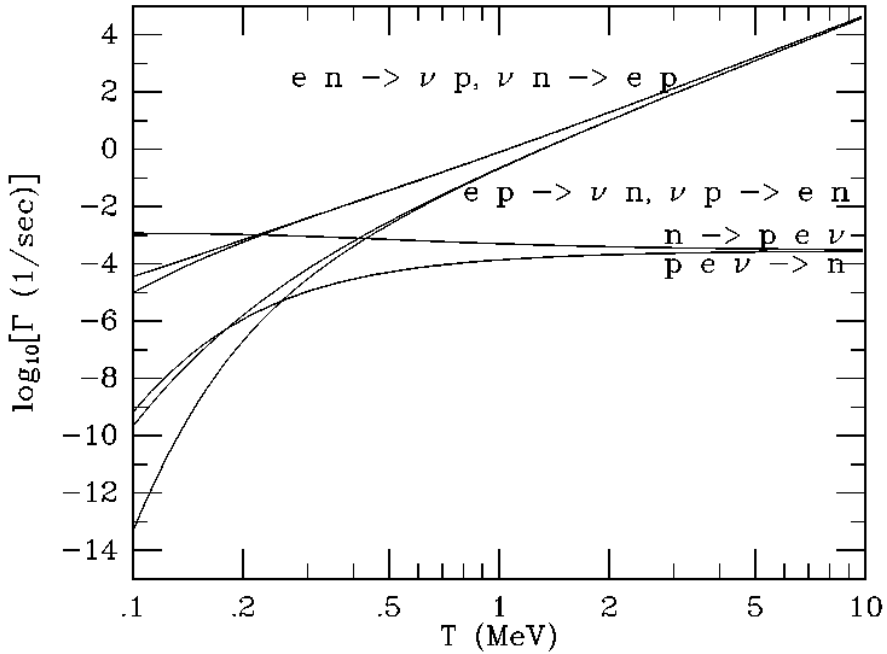,width=5in}}
\caption{Weak rates as a function of temperature (infinite-nucleon-mass limit).
Note, freeze-out of the $n/p$ ratio occurs at $T \simeq 0.8\,{\rm MeV}$
and nucleosynthesis begins in earnest at $T \simeq 0.1\,{\rm MeV}$.}
\label{fg:rates}
\end{figure}

The weak interactions that interconvert neutrons and protons, $n\leftrightarrow p + e
+\nu$, $n+e\leftrightarrow p+\nu$, and $n+\nu \leftrightarrow p+e$, play a crucial
role as they govern the neutron fraction, and the neutron fraction ultimately
determines the amount of nucleosynthesis that takes place.  (Here and throughout we
use $e$ to indicate electron or positron, and $\nu$ to indicate electron neutrino or
antineutrino; the appropriate particle or antiparticle designation follows from
charge and lepton-number conservation.) For reference, Fig.~\ref{fg:rates} shows
the weak rates as a function of temperature.

\subsection{Rate Expressions}
Generally, the weak-interaction rates may be expressed as integrals over
phase space. For definiteness, consider the
process $ep\rightarrow\nu n$. The other processes involve similar
expressions. The rate, reactions per incident nucleon per time, is given by a
twelve-dimensional integral,
\begin{equation}
\Gamma_{ep\leftrightarrow\nu n} = {1\over n_p} \int\prod_i{d\Pi_i} (2\pi)^4 
\left|{\cal M}\right|^2 \delta^{(4)}(e+p-\nu-n) 
f_e f_p (1-f_\nu) (1-f_n), \label{eq:12D}
\end{equation}
where $n_p$ is the number density of the incident nucleon, $d\Pi_i = d^3p_i /
[(2\pi)^32E_i]$ is the Lorentz invariant phase-space element, $\left|{\cal
M}\right|^2$ is the weak-interaction matrix element summed over initial and
final spins (see Appendix~A), $e$, $p$, $\nu$ and $n$ are the four-momenta of
the particles, and the delta function $\delta^{(4)}(e+p-\nu-n)$ expresses
conservation of four-momentum. In the rest frame of the thermal radiation,
the phase-space densities $f_e$, $f_p$, $f_\nu$ and $f_n$ are given by the
usual Fermi-Dirac distribution functions.

Assuming only the isotropy of space, we can integrate over $d\Omega_e$, the
direction of the electron's three-momentum and $d\phi_p$, the azimuthal angle
of the proton's three-momentum. After applying conservation of three-momentum
to eliminate $d^3 p_n$ and conservation of energy to eliminate $d p_\nu$ the
expression simplifies to the following five-dimensional integral:
\begin{eqnarray}
\Gamma_{ep\leftrightarrow\nu n} 
& = & 
{1 \over 2^9\pi^6 n_p}
\int dp_e dp_p d\cos\theta_p d\cos\theta_\nu d\phi_\nu
\nonumber \\
&   &
\times{p_e^2 p_p^2 E_\nu\over E_e E_p E_n} {1\over {\cal J}} 
\left|{\cal M}\right|^2 f_e f_p (1-f_\nu) (1-f_n),  \label{eq:5D} \\
{\cal J} & = & 
1 + { E_\nu \over E_n }
\left( 1 - { ({\bf p}_e+{\bf p}_p) \cdot {\bf p_\nu} \over E_\nu^2 } \right) ,
\label{eq:jacob}
\end{eqnarray}
where $E_e$, $E_p$, $E_\nu$, and $E_n$ denote the energies of the respective
particles and $\cal J$ is the Jacobian introduced in integrating the energy
part of the delta function.  We take ${\bf p_e}$ parallel to the polar
axis ($\theta_e = \phi_e = 0$) and $\phi_p = 0$.
We need an expression for $E_\nu$
in terms of the integration variables $p_e, p_p, \theta_p, \theta_\nu$, and
$\phi_\nu$; it is given by
\begin{eqnarray}
p_\nu & = & {A^2 B + 2E \sqrt{A^4-m_\nu^2(4E^2-B^2)} \over 4E^2-B^2} ,
\nonumber \\
A^2 & \equiv & 2E_e E_p + 
               m_\nu^2 - m_n^2 - m_e^2 - m_p^2 - 2 p_e p_p \cos\theta_p,
\nonumber \\
B & \equiv & 2\left[ p_e \cos\theta_\nu + p_p\left(\cos\theta_p \cos\theta_\nu
        + \sin\theta_p \sin\theta_\nu \cos\phi_\nu \right) \right] .
\end{eqnarray}
where $E=E_e+E_p$.  If $m_\nu$ is taken to be zero (or very small)
these equations simplify further.   We left them in
this form because for two-body processes in which the electron is
the outgoing lepton, $m_\nu \rightarrow m_e \neq 0$.

\subsection{Infinite-Mass Approximation}

If the nucleons are assumed to be infinitely massive,\footnote{More precisely, $m_n,
m_p \rightarrow \infty$ with $m_n-m_p = Q$ fixed.} two things happen. The kinematics
simplifies because the kinetic energy of the nucleons can be neglected
($E_e-E_\nu = Q$), and the matrix element simplifies, $\left|{\cal M}\right|^2 =
2G_F^2(1+3g_A^2)2E_e 2E_p 2E_\nu 2E_n$. Here $G_F = 1.166 \times 10^{-5} {\rm
GeV}^{-2}$ is the Fermi constant and $g_A \simeq 1.257$ is the ratio of axial to
vector coupling to the nucleon. The rate expression now reduces to the familiar
form~\cite{Weinberg72}:
\begin{equation}
\Gamma_{ep\leftrightarrow\nu n}^\infty = {G_F^2(1+3g_A^2)m_e^5 \over2\pi^3}
\int_q^\infty {\epsilon (\epsilon^2-q^2)^{1/2}\over
\left[1+\exp(\epsilon z)\right]\left[1+\exp((q-e)z_\nu))\right]} ,
\label{eq:1D}
\end{equation}
where $T$ is the photon temperature, $T_\nu$ is the neutrino temperature,
$\epsilon\equiv E_e/m_e$, $z\equiv m_e/T$, and $z_\nu\equiv m_e/T_\nu$. Expressions
for the other five $n\leftrightarrow p$ processes are similar. This is the
approximation used in the standard nucleosynthesis code.

In the era preceding nucleosynthesis, when to a good approximation baryons exist as
free neutrons and protons, the neutron fraction $X_n$ is governed by
\begin{equation}
\dot{X}_n = -X_n \Gamma_{n \rightarrow p} + (1-X_n)\Gamma_{p \rightarrow n},
\end{equation}
where $X_p = 1-X_n$, and
\begin{eqnarray}
\Gamma_{p\rightarrow n} & \equiv & 
\Gamma_{e p\rightarrow \nu n} + \Gamma_{\nu p\rightarrow e n} + 
  \Gamma_{p e\nu \rightarrow n} \nonumber \\
\Gamma_{n\rightarrow p} & \equiv & 
\Gamma_{e n\rightarrow \nu p} + \Gamma_{\nu n\rightarrow e p} + 
  \Gamma_{n \rightarrow p e \nu} .
\end{eqnarray}

In thermal equilibrium, which holds when the rates are much greater than the
expansion rate of the Universe,
\begin{equation}
\left( {n\over p} \right) = 
{\Gamma_{p\rightarrow n} \over \Gamma_{n\rightarrow p}} =
\left({m_n\over m_p}\right)^{3/2} e^{-Q/T} .
\label{eq:np}
\end{equation}
In the infinite-mass limit, $m_n/m_p = 1$, and
\begin{equation}
\left( {n \over p} \right)^\infty = 
{\Gamma_{p\rightarrow n}^\infty \over \Gamma_{n\rightarrow p}^\infty} = e^{-Q/T}.
\label{eq:npInf}
\end{equation}
The second equality in both equations is just detailed balance, which is one of the
checks that we will use to test our results.

\section{Finite-Mass Corrections}

We carried out the five-dimensional integration using the Monte Carlo technique. The
challenge was to compute a finite-mass correction, whose size is of order a few
percent, to a relative accuracy of a few percent. Hence, we had to evaluate the rate
integral to order 0.1\% accuracy. The statistical uncertainty in integrating
the rate expressions was typically around 0.3\% for several minutes of
computer time. Since this scales as $1/\sqrt{N}$, where $N$ is the number of
function evaluations, we could have achieved the accuracy requirements by
simply increasing the integration time by a factor of ten or so. However, we
found a more efficient way of proceeding, which also allowed
other checks to be made.

\begin{figure}[t]
\centerline{\psfig{figure=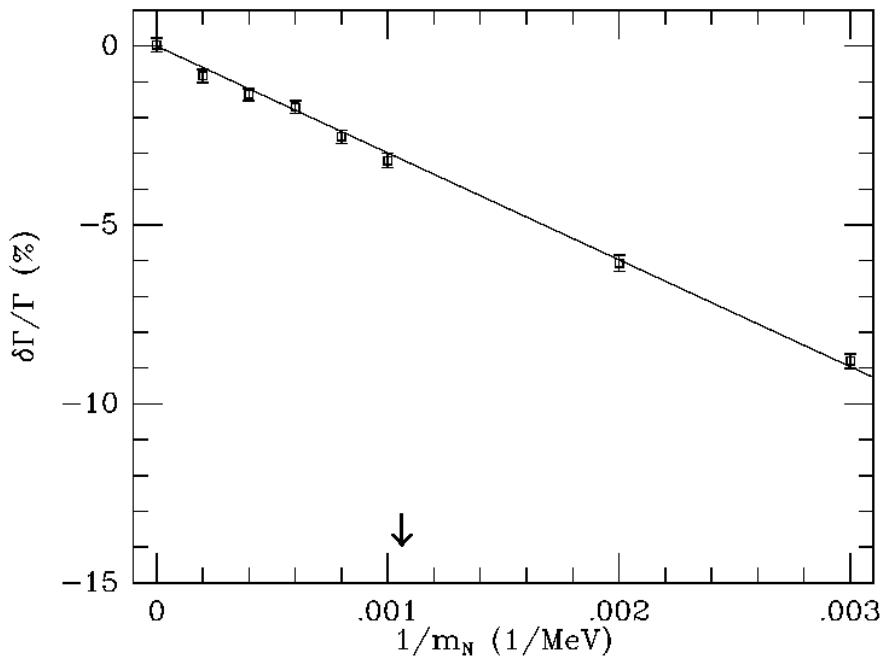,width=5in}}
\caption{Finite-nucleon-mass correction to $\nu p \rightarrow e n$ at
$T=1.0\,{\rm MeV}$ as a function of $1/m_N$. The error bars represent the
estimated statistical uncertainty. The arrow indicates the actual value of the
nucleon mass, $m_N \simeq 940\,\MeV$. Note, the correction vanishes for $m_N
\rightarrow \infty$ and is linear in $1/m_N$.}
\label{fg:p3t1}
\end{figure}

For $E / m_N \ll 1$ the finite-mass correction to the weak rates should vary
linearly with $1/m_N$ ($E \sim 1\,$MeV is the characteristic energy scale: $E
\sim Q, m_e, T$.)  As $E/m_N \rightarrow 0$, the correction must approach
zero. We adopted the following strategy. We performed a series of runs where
we fixed the temperature but varied the nucleon mass. We found the finite-mass
correction ($ \equiv \delta\Gamma = \Gamma - \Gamma^\infty$) by applying a
linear fit to the data and interpolating to $m_N \simeq
940\,\MeV$. Figure~\ref{fg:p3t1} shows the result of this procedure for the
process $ep \rightarrow \nu n$. The linearity of $\delta\Gamma / \Gamma$ in
$1/m_N$ is manifest as is the intercept at $\delta\Gamma / \Gamma = 0$ for
$m_N \rightarrow \infty$.

\begin{figure}[t]
\centerline{\psfig{figure=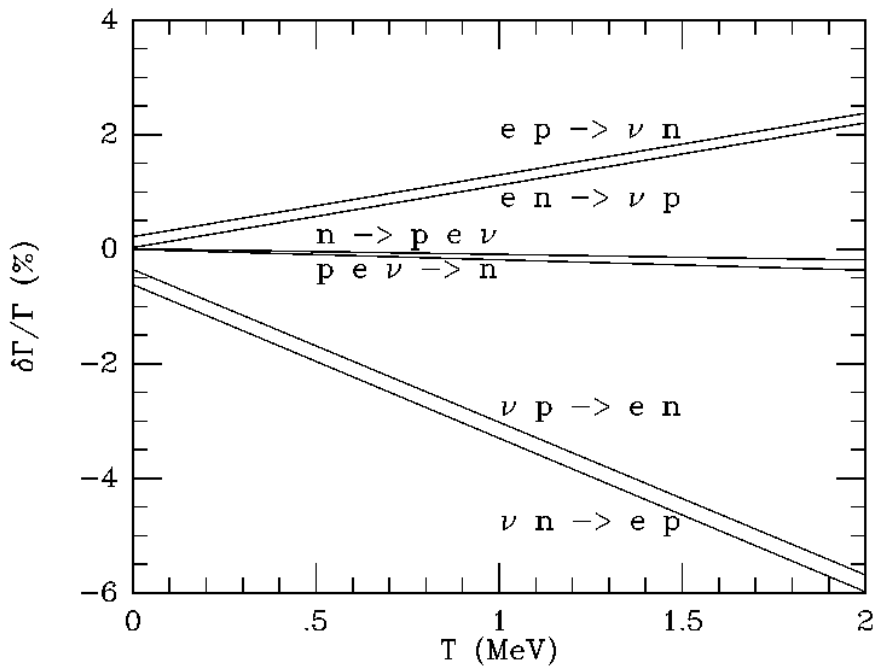,width=5in}}
\caption{Finite-nucleon-mass corrections to the weak rates as a function of
temperature.}
\label{fg:finmass}
\end{figure}

We found that the finite-mass correction for each process varies linearly with
temperature. Therefore, we applied a linear fit to the correction as a
function of temperature. In the end we were able to achieve the required order
$0.1 \%$ absolute accuracy in the rate integrals. Figure~\ref{fg:finmass}
shows the finite-mass corrections $\delta\Gamma / \Gamma$ for all six
processes that interconvert neutrons and protons. The corrections are, as
expected, of order a few percent.

\subsection{Tests of finite-mass corrections}
We applied several checks to our results. The simplest was to let $1/m_N$
vary. As $1/m_N \rightarrow 0$ the finite-mass rate corrections should approach
zero; Fig.~\ref{fg:p3t1} illustrates that they do. While Fig.~\ref{fg:p3t1}
illustrates convergence for a particular process and
temperature, convergence was observed for all processes and temperatures. 

\begin{figure}[t]
\centerline{\psfig{figure=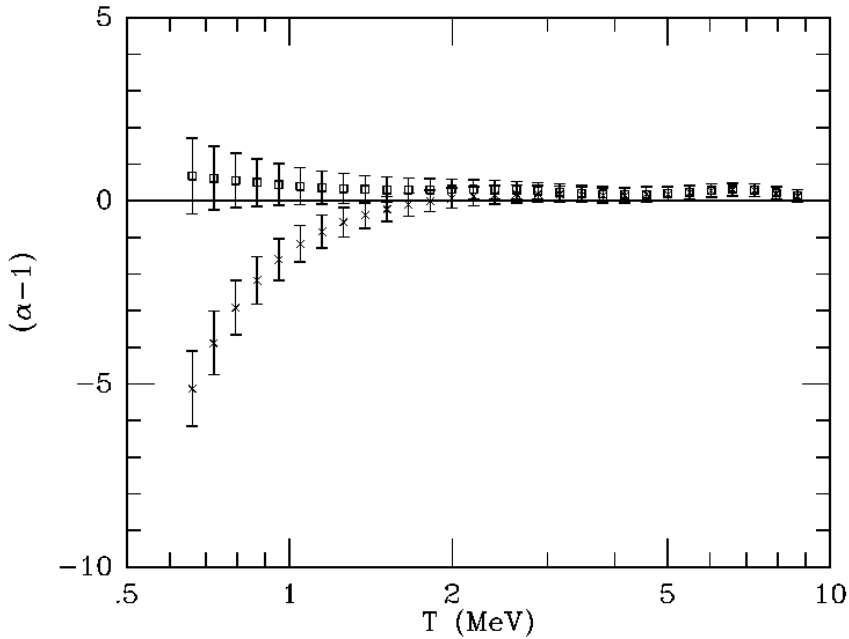,width=5in}}
\caption{Detailed-balance test. For $\alpha = 1$ detailed balance is satisfied (see
text). For the boxes, $T_\nu = T_\gamma$ is enforced; for the crosses
neutrinos are assumed to be decoupled, so that $T_\nu < T_\gamma$ for
$T_\gamma \protect\la 1 {\rm MeV}$. Since the finite-mass corrections are
obtained from fits to $1/m_N$ and $T_\gamma$, the errors are strongly
correlated.}
\label{fg:dbal}
\end{figure}

Another test is provided by detailed balance. Detailed balance requires that
\begin{equation}
\left({ \Gamma_{p\rightarrow n} \over \Gamma_{n\rightarrow p} } \right) =
\left( {n\over p} \right)_{\rm EQ} =
\left({m_n \over m_p}\right)^{3/2} e^{-Q/T} \simeq 
\left({3\over 2}{Q\over m_N}+1\right) e^{-Q/T} ,
\end{equation}
while the neutrinos are coupled to the photon plasma and $T_\nu = T_\gamma$. We
numerically determined $\Gamma_{n\rightarrow p} / \Gamma_{p\rightarrow n}$, and
expressed it as
\begin{equation}
{ \Gamma_{p\rightarrow n} \over \Gamma_{n\rightarrow p} } \simeq 
\left({3\over 2}{Q \over m_N}\alpha+1\right) e^{-Q/T} ;
\end{equation}
$\alpha = 1$ corresponds to detailed balance being satisfied. Figure~\ref{fg:dbal} shows
that detailed balance is satisfied while the neutrinos are coupled to
the photons and also when the neutrino temperature is set equal to the photon
temperature.

Finally, we were able to derive an independent and simpler
expression for the finite-mass rates
but using the infinite-mass matrix element. This provides a check of the complicated
kinematics that arise for $m_N \neq \infty$. In Eq.~(\ref{eq:12D}) the
three-momentum part of the delta function is expanded in complex exponentials,
\begin{eqnarray}
\Gamma_{ep\rightarrow\nu n} & = & {1\over n_p} \int\prod_i{d\Pi_i} d^3 x (2\pi)
e^{i {\bf x} \cdot ({\bf p}_e + {\bf p}_p - {\bf p}_\nu - {\bf p}_n)}
\left|{\cal M}\right|^2  \nonumber \\
 & & \times \delta(E_e+E_p-E_\nu-E_n) f_e f_p (1-f_\nu) (1-f_n).
\end{eqnarray}
If the integrand has no angular dependence, as is the case for the infinite-mass
matrix element, the expression can be reduced to three-dimensional integral.
(With
angular dependence, the reduction can be done term by term in the angular expansion
of the matrix element, but the resulting expression becomes very complicated.)
For the infinite-mass matrix element all of the angular integrals can be done,
\begin{eqnarray}
\Gamma^{3}_{ep\rightarrow\nu n} & = & 
{G_F^2 \left(1+3g_A^2\right) \over \pi^6 n_p} \int \prod_i dp_i\ p_i \sin(x p_i)
{d^3 x \over x^2} \nonumber \\
& & \times \delta(E_e + E_p - E_\nu - E_n) f_e f_p (1-f_\nu) (1-f_n) .
\end{eqnarray}
The integral over $dx$ can be done by a standard contour integration,
\begin{eqnarray}
\Gamma^{3}_{ep\rightarrow\nu n} & = & -{G_F^2(1+3g_A^2)\over 2^4\pi^5 n_p} 
\int dE_e dE_p dE_\nu E_e E_p E_\nu (E_p+E_e-E_\nu)
\nonumber \\
&   & \times (\sum_{i=1}^8 t_i) f_e f_p (1-f_\nu) (1-f_n) ,
\label{eq:3D}
\end{eqnarray}
where the energy delta function has been used to carry out the $dE_n$
integral and
\begin{eqnarray}
\sum_{i=1}^8 t_i  & = &
\phantom{+} \left|p_e+p_p+p_\nu+p_n\right| + \left|p_e+p_p-p_\nu-p_n\right| \nonumber \\
& & 
+ \left|p_e-p_p+p_\nu-p_n\right| + \left|p_e-p_p-p_\nu+p_n\right| \nonumber \\
& & 
- \left|p_e+p_p+p_\nu-p_n\right| - \left|p_e+p_p-p_\nu+p_n\right| \nonumber \\
& & 
- \left|p_e-p_p+p_\nu+p_n\right| - \left|p_e-p_p-p_\nu-p_n\right| .
\end{eqnarray}
Using standard techniques we carried out the numerical integration.

\begin{figure}[t]
\centerline{\psfig{figure=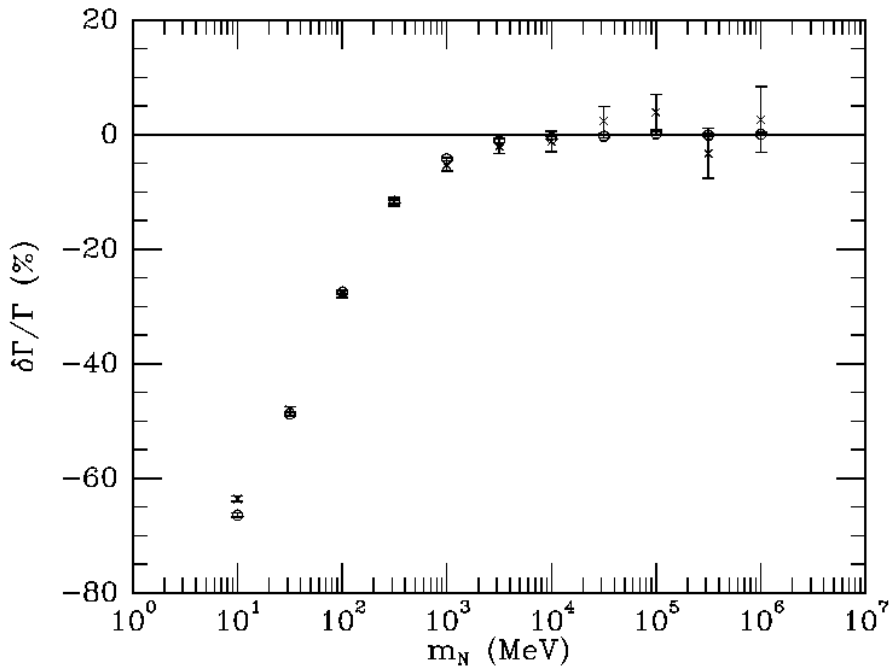,width=5in}}
\caption{Kinematics test for $T=5.0 {\rm MeV}$ (see text). Circles are the
five-dimensional integration and crosses are the three-dimensional integration.}
\label{fg:oneT}
\end{figure}

\begin{figure}[t]
\centerline{\psfig{figure=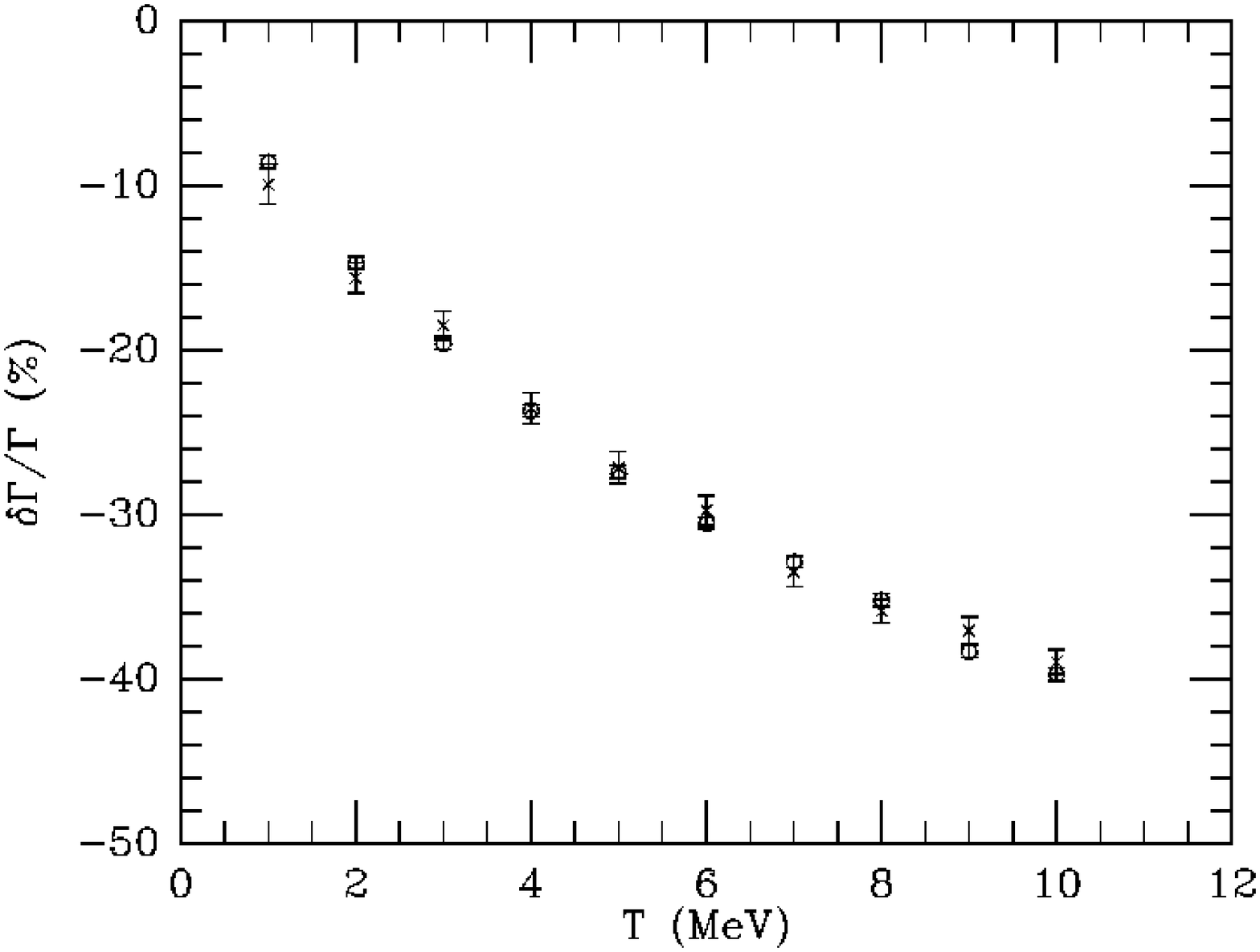,width=5in}}
\caption{Kinematics test with $m_N = 100.0 {\rm MeV}$ (see text). Circles are the
five-dimensional integration and crosses are the three-dimensional integration.}
\label{fg:oneM}
\end{figure}

For our test we compare $\Gamma^{3}$ with the rate obtained by inserting the
infinite-mass matrix element into Eq.~(\ref{eq:5D}). This rate, denoted by
$\Gamma^{5}$, is given by
\begin{eqnarray}
\Gamma_{ep\rightarrow\nu n}^{5}
& = & 
{G_F^2 \left( 1+3g_A^2 \right) \over 2^4\pi^6 n_p}
\int dp_e dp_p d\cos\theta_p d\cos\theta_\nu d\phi_\nu
\nonumber \\
&   & 
\times p_e^2 p_p^2 p_\nu E_\nu  {1\over {\cal J}} 
f_e f_p (1-f_\nu) (1-f_n).
\end{eqnarray}
Formally, $\Gamma^{3}$ and
$\Gamma^{5}$ are identical. Numerically, they are computed using independent
techniques. Comparing them provides a stringent test of
kinematics. Figure~\ref{fg:oneT} shows this comparison as a function of $m_N$ with $T
= 5\:{\rm MeV}$. Figure~\ref{fg:oneM} shows the comparison as a function of $T$ with
$m_N = 100\:{\rm MeV}$.  In both cases, $\Gamma^3$ and $\Gamma^5$ agree within
estimated numerical uncertainties. In conclusion, because our finite-mass rate
corrections pass these three important tests, we have confidence that they are
correct and accurate.

\section{Changes to the Standard Code}

\begin{figure}
\centerline{\psfig{figure=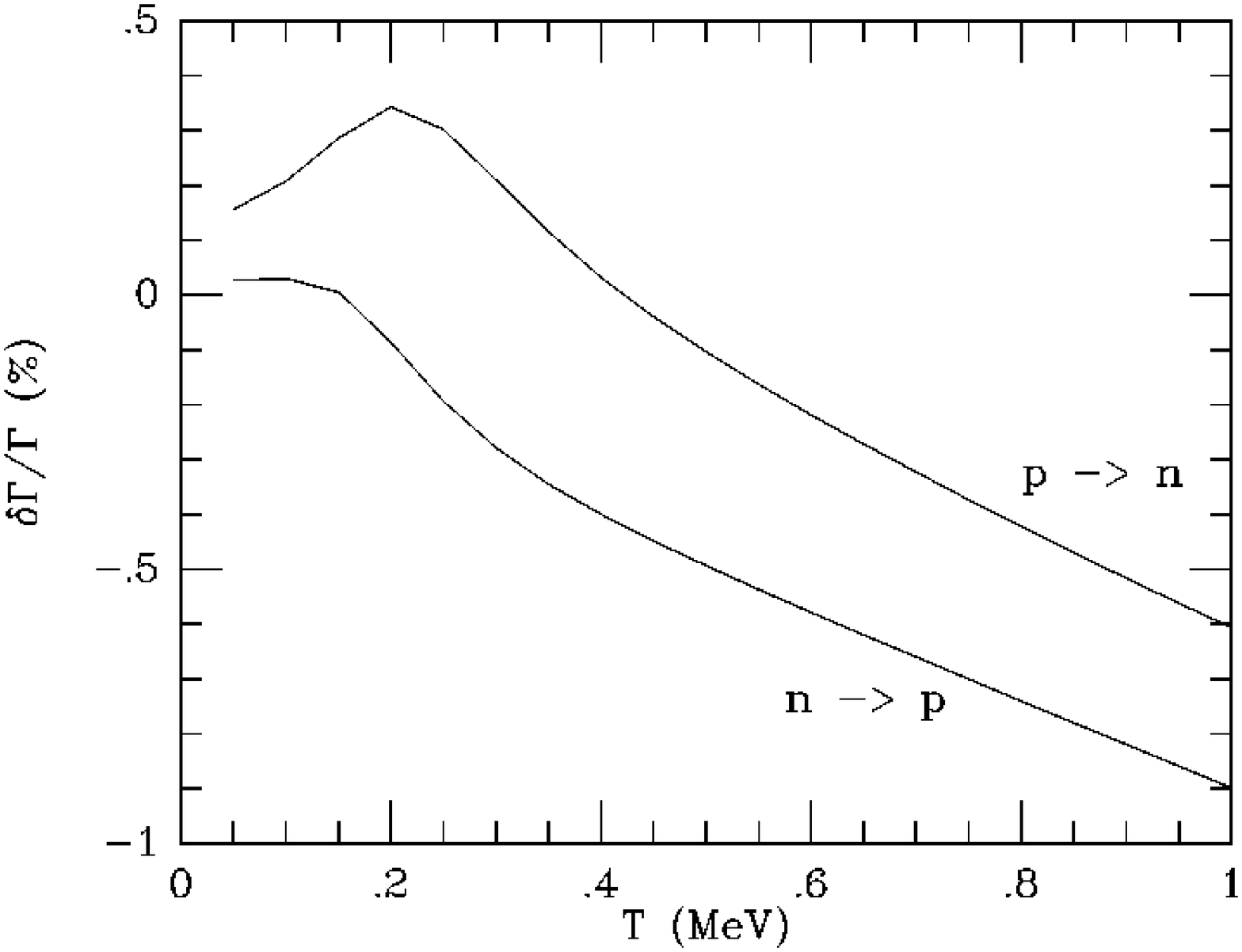,width=5in}}
\caption{Corrections to $\Gamma_{p \rightarrow n}(T)$ and $\Gamma_{n \rightarrow
p}(T)$.}
\label{fg:deltarate}
\end{figure}

The standard code calculates the weak rates at each temperature step. (The user is
offered the choice of either integrating the rates of using fits to them. We chose to
have the code integrate the rates at each step.) The rates are obtained by
integrating equations like Eq.~(\ref{eq:1D}), which assume that the nucleons are
infinitely massive. Our approach was to implement the finite-mass rate corrections as
a multiplicative factor at each temperature step:
\begin{eqnarray}
\Gamma_{p \rightarrow n}  & = & \Gamma_{p \rightarrow n}^\infty 
\left( 1 + {\delta\Gamma \over \Gamma} \right) 
\label{eqn:codeInputA} \nonumber  \\
\Gamma_{n \rightarrow p} & = & \Gamma_{n \rightarrow p}^\infty 
\left( 1 + {\delta\Gamma \over \Gamma} \right) ,
\label{eqn:codeInputB}
\end{eqnarray}
where $\delta\Gamma = \sum_i \delta\Gamma_i, \Gamma = \sum_i \Gamma_i$ and $i$ runs over
the three reactions that convert neutrons to protons (or vice versa). A plot of the
relative finite-mass rate corrections as a function of temperature is shown in
Fig.~\ref{fg:deltarate}. At the crucial freeze-out epoch ($T \simeq 0.8\MeV$), both
rates are corrected downward, with $\delta\Gamma_{p\rightarrow n} /
\Gamma_{p\rightarrow n} = -0.43\%$ and $\delta\Gamma_{n\rightarrow p} /
\Gamma_{n\rightarrow p} = -0.55\%$

One subtlety that we have not mentioned is the fact that in the standard code the weak
rates are normalized to the measured free-neutron decay rate. This means that
free-neutron decay has already been ``corrected'' for the zero-temperature,
finite-nucleon-mass effect. Since the other rates are normalized to $\tau_n$, they
too include a piece of the finite-nucleon-mass correction, where the
finite-nucleon-mass correction to free-neutron decay is $\delta_n$. To properly implement our
results in the standard code we must ``remove'' this correction, so that
Eq.~(\ref{eqn:codeInputA}) is actually implemented as
\begin{eqnarray}
\Gamma_{p \rightarrow n} = \Gamma_{p \rightarrow n}^{\rm std} 
\left( 1 + {\delta\Gamma \over \Gamma} \right) \left( 1 - \delta_n \right) 
\nonumber \\
\Gamma_{n \rightarrow p} = \Gamma_{n \rightarrow p}^{\rm std}
\left( 1 + {\delta\Gamma \over \Gamma} \right) \left( 1 - \delta_n \right).
\end{eqnarray}

To compute $\delta_n$ we began with the twelve-dimensional phase-space
integral for neutron decay,
\begin{equation}
\Gamma_{n \rightarrow p e \nu} = {1\over n_p} \int\prod_i{d\Pi_i} (2\pi)^4
\left|{\cal M}\right|^2 \delta^{(4)}(e+p-\nu-n) 
f_n (1-f_p) (1-f_e) (1-f_\nu) .
\end{equation}
In the zero-temperature limit, $f_n = (n_n/2)\:(2\pi)^3\:\delta^{(3)}({\bf p_n})$,
and $(1-f_p) = (1-f_e) = (1-f_\nu) = 1$. This expression can be simplified to the
following two-dimensional integral:
\begin{eqnarray}
\Gamma_{n \rightarrow p e \nu} & = &
{1 \over 2^7 \pi^3} \int_{m_e}^{Q- (Q^2 - m_e^2)/2m_N} dE_e 
\int_{-1}^1 d\cos\theta_\nu
{ \left|{\cal M}\right|^2 p_e E_\nu \over m_n E_p \left|{\cal J}\right|}, \\
\left|{\cal J}\right| & = & 
1 + {E_\nu + p_e \cos\theta_\nu \over E_p} , \\
E_p & = & m_n - E_\nu - E_e, \\
E_\nu & = &
{ m_n^2 - m_p^2 + m_e^2 - 2m_n E_e \over
 2 \left( m_n - E_e + p_e \cos\theta_\nu \right) } ,
\end{eqnarray}
where $\left|{\cal J}\right|$ is the Jacobian introduced due to integration over the
energy delta function. Note that the full matrix element is used. Using standard
techniques, we integrated this expression and found the zero-temperature neutron
decay correction, $\delta_n \equiv \Gamma_{n\rightarrow pe\nu}
-\Gamma^\infty_{n\rightarrow pe\nu}= -0.206$\%.

\section{Results and Conclusions}
\begin{figure}[t]
\centerline{\psfig{figure=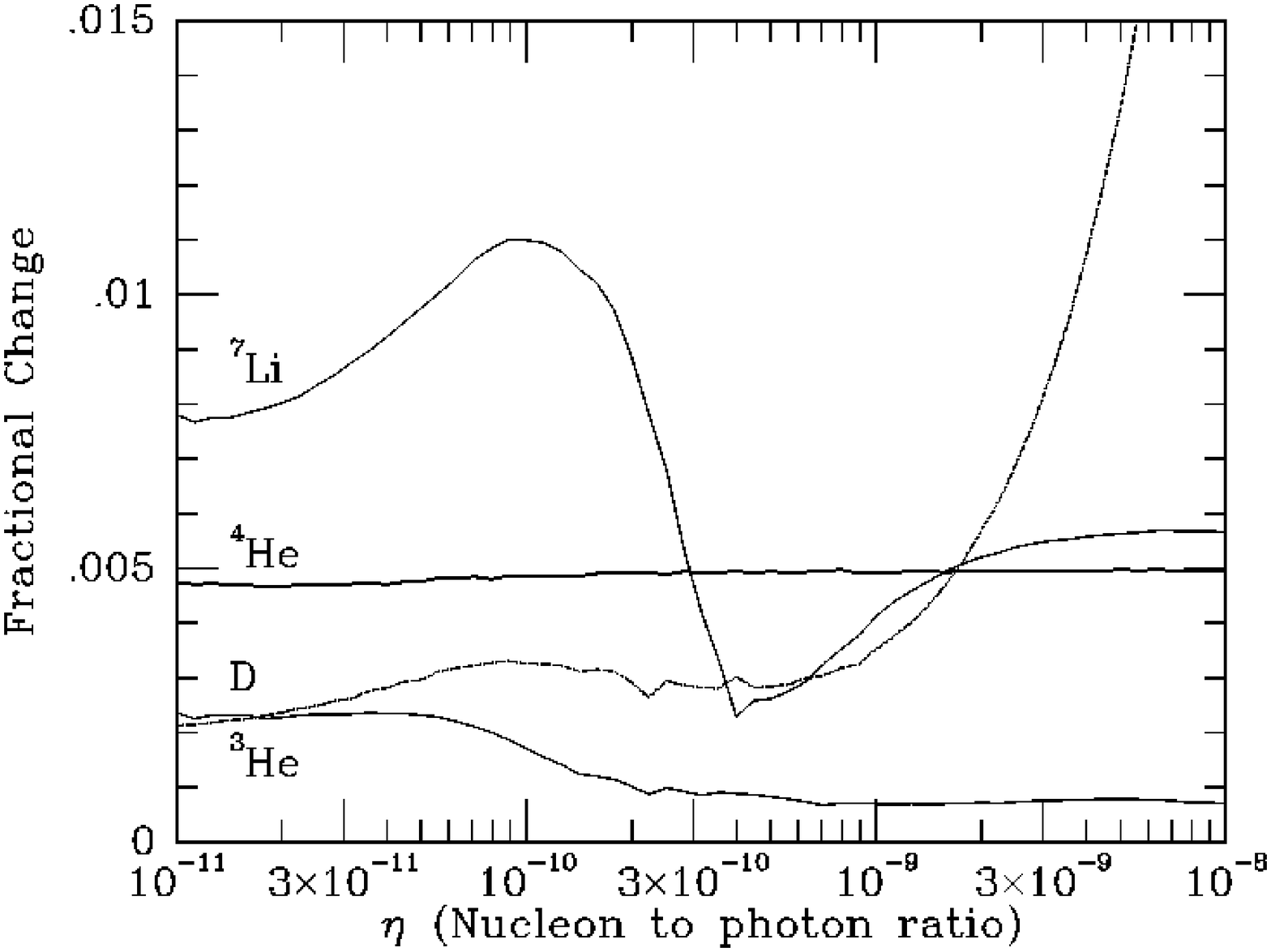,width=5in}}
\caption{Finite-mass corrections to light element abundances. For $^4$He the
fractional change is in the mass fraction; for the others it is in the number
relative to H.}
\label{fg:fract}
\end{figure}

Our results, which were obtained for three massless-neutrino species and a mean
neutron lifetime of 887 sec \cite{PDG96},
are shown in Fig.~\ref{fg:fract}. Plotted are the
fractional changes in the mass abundances of D, $^3\rm He$, $^4\rm He$ and $^7\rm Li$
versus the baryon-to-photon ratio $\eta$. The relative change in $^4\rm He$ is
approximately 0.47\% -- 0.50\% over a large interval, $10^{-11}\leq\eta\leq
10^{-8}$. Over the most interesting part of the interval, $\eta = (2 - 7) \times
10^{-10}$, $\Delta Y/Y = (0.49 - 0.50)\%$. Over the full interval in $\eta$, the
fractional change in the other light elements span the range 0.06\% to 2.8\%.  Unlike
$^4\rm He$, the inferred primordial abundances of these elements are known to nowhere
near this accuracy and these changes are of little relevance at present.

Several years ago the effect of finite nucleon mass on the yields of nucleosynthesis
was estimated by Gyuk and Turner~\cite{Gyuk93} based on an approximation scheme
developed by Seckel~\cite{Seckel93}. Seckel estimated the lowest-order correction,
i.e., terms of order $1/m_N$, to the weak rates and Gyuk and Turner incorporated them
into the standard code. Our finite-nucleon-mass corrections are generally consistent
with Seckel's estimates for the changes in the rates; however,
there are significant
qualitative and quantitative differences. Likewise, our calculation of $\Delta Y$ is
consistent, but significantly smaller, than that of Gyuk and Turner who found
$\Delta Y \simeq 0.006 Y$ based on Seckel's scheme.

\section*{Acknowledgements}

We would like to thank Lars Bergstrom for checking the matrix element, Sasha Dolgov
for suggesting the trick that allows the rate integral to be reduced to a
three-dimensional integral, and David Seckel for many helpful conversations and
comparisons to his earlier work. This work was supported in part by the DOE (at
Chicago and Fermilab) and by NASA at Fermilab through grant NAG~5--2788.

\appendix
\section{Appendix: The Matrix Element}

Here we present the full matrix element to leading order in the weak
interaction coupling constant~\cite{Seckel93,Bergstrom96}. The particles
labeled by 1 and 3 are the incident and outgoing leptons, and by 2 and 4 are
the incident and outgoing nucleons. The matrix element is expressed in terms
of the relativistic invariants $s$ and $t$. Here, $ s =\left(1+2\right)^2 =
\left(3+4\right)^2
$ 
and 
$ t = \left(1-3\right)^2 =
\left(2-4\right)^2 $ , 
where $1,2,3,4$ are the four-momenta of the respective particles, $f_2 =
1.81$ is the anomalous weak magnetic moment of the nucleon, and $g_A = 1.257$
is the ratio of axial to vector coupling of the nucleon.

This expression applies to all six weak processes. For anti-leptonic, two-body
processes, e.g., $e^+ n
\rightarrow \bar{\nu}_e p$, $g_A \rightarrow -g_A$. For $n\rightarrow p e \nu$, $s =
\left(-1+2\right)^2$ and $t = \left(-1-3\right)^2$ and for $p e \nu\rightarrow n$,
$s =
\left(1+2\right)^2$ and $t = \left(1+3\right)^2$.

\begin{eqnarray*}
\left|{\cal M}\right|^2 &=& G_F^2 \left(t_1 + t_2 + t_3 + t_4 + t_5 + t_6 \right), \\
t_1 &=& 16 f_2 /m_N \times \\
&& 
  (	
	-m_1^4 m_4 +
	m_1^2 m_2^3 - 
	m_1^2 m_2^2 m_4 +
	m_1^2 m_2 m_3^2 - 
	m_1^2 m_2 s  \nonumber \\
&&
	- m_1^2 m_2 t +
	m_1^2 m_3^2 m_4 + 
	m_1^2 m_4 s - 
	m_2^3 t + 
	m_2^2 m_4 t  \nonumber \\ 
&&
	- m_2 m_3^4 -
	m_2 m_3^2 m_4^2 + 
	m_2 m_3^2 s + 
	m_2 m_4^2 t + 
	m_2 t^2  \nonumber \\ &&
	+ m_3^2 m_4^3 - 
	m_3^2 m_4 s - 
	m_3^2 m_4 t - 
	m_4^3 t + 
	m_4 t^2
  ) , \\
t_2 &=& 8  \times \\
&&
  (
	m_1^2 m_2^2 - 
	2m_1^2 m_2 m_4 +
	2 m_1^2 m_3^2 + 
	m_1^2 m_4^2 -
	2m_1^2 s \nonumber \\ &&
	- m_1^2 t + 
	m_2^2 m_3^2 +
	2m_2^2 m_4^2 -
	2m_2^2 s - 
	m_2^2 t  \nonumber \\ &&
	- 2m_2 m_3^2 m_4 + 
	2 m_2 m_4 t + 
	m_3^2 m_4^2 -2m_3^2 s - 
	m_3^2 t - 
	2m_4^2 s \nonumber \\ &&
	- m_4^2 t + 
	2s^2 + 
	2st + 
	t^2
   ) , \\
t_3 &=& 4 f_2^2 / m_N^2 \times \\
&&
  (
	m_1^4 m_2^2 -
	2 m_1^4 m_2 m_4 -
	3 m_1^4 m_4^2 -
	m_1^4 t +
	2 m_1^2 m_2^4 \nonumber \\ &&
	+ 2 m_1^2 m_2^2 m_3^2 -
	4 m_1^2 m_2^2 s -
	3 m_1^2 m_2^2 t +
	4 m_1^2 m_2 m_3^2 m_4 -
	2 m_1^2 m_2 m_4 t \nonumber \\ &&
	+ 2 m_1^2 m_3^2 m_4^2 -
	2 m_1^2 m_3^2 t -
	2 m_1^2 m_4^4 +
	4 m_1^2 m_4^2 s +
	m_1^2 m_4^2 t \nonumber \\ &&
	+ 4 m_1^2 s t +
	m_1^2 t^2 -
	2 m_2^4 m_3^2 -
	2 m_2^4 t -
	3 m_2^3 m_3^4 \nonumber \\ &&
	+ 4 m_2^2 m_3^2 s +
	m_2^2 m_3^2 t +
	m m_2^2 s t +
	2 m_2^2 t^2 - 
	2 m_2 m_3^4 m_4 \nonumber \\ &&
	- 2 m_2 m_3^2 m_4 t +
	4 m_2 m_4 t^2 +
	m_3^4 m_4^2 -
	m_3^4 t +
	4 m_3^2 s t \nonumber \\ &&
	+m_3^2 t^2 -
	2 m_4^4 t +
	4 m_4^2 s t +
	2 m_4^2 t^2 -
	4 s^2 t \nonumber \\ &&
	- 4 s t^2
  ) , \\
t_4 &=& 8 g_A^2 \times \\
&&
  (
	m_1^2 m_2^2 +
	2 m_1^2 m_2 m_4 +
	2 m_1^2 m_3^2 +
	m_1^2 m_4^2 -
	2 m_1^2 s \nonumber \\ &&
	- m_1^2 t +
	m_2^2 m_3^2 +
	2 m_2^2 m_4^2 -
	2 m_2^2 s -
	m_2^2 t \nonumber \\ &&
	+ 2 m_2 m_2^2 m_4 -
	2 m_2 m_4 t +
	m_3^2 m_4^2 -
	2 m_3^2 s -
	m_3^2 t \nonumber \\ &&
	- 2 m_4^2 s  -
	m_4^2 t +
	2 s^2 +
	2 s t +
	t^2
  ) , \\
t_5 &=& 16 g_A^2 f_2 / m_2 \times \\
&&
  (
	-m_1^2 m_2^3 -
	m_1^2 m_2^2 m_4 +
	m_1^2 m_2 m_4^2 +
	m_1^2 m_2 t +
	m_1^2 m_4^3 \nonumber \\ &&
	+ m_1^2 m_4 t +
	m_2^3 m_3^2 +
	m_2^3 t +
	m_2^2 m_3^2 m_4 +
	m_2^2 m_4 t \nonumber \\ &&
	- m_2 m_3^2 m_4^2 +
	m_2 m_3^2 t +
	m_2 m_4^2 t -
	2 m_2 s t -
	m_2 t^2  \nonumber \\ &&
	- m_3^2 m_4^3 +
	m_3^2 m_4 t +
	m_4^3 t -
	2 m_4 s t - 
	m_4 t^2
  ) , \\
t_6 &=& 16 g_A  \times \\
&&
  (
	-m_1^2 m_2^2 +
	m_1^2 m_4^2 +
	m_1^2 t +
	m_2^2 m_3^2 +
	m_2^2 t \nonumber \\ &&
	- m_3^2 m_4^2 +
	m_3^2 t +
	m_4^2 t -
	2 s t -
	t^2
  ).
\end{eqnarray*}


\newpage

\end{document}